\documentclass[iicol]{sn-jnl}

\usepackage{subcaption}
\usepackage{amsmath,amssymb,amsfonts}
\usepackage{braket}
\usepackage{xcolor}
\usepackage{hyperref}
\usepackage{appendix}
\usepackage[font=footnotesize]{caption}
\usepackage{subcaption}

\usepackage[style=numeric-comp,backend=biber,sorting=none]{biblatex}
\DeclareSourcemap{
  \maps[datatype=bibtex]{
    \map[overwrite]{
      \step[fieldsource=doi, final]
      \step[fieldset=url, null]
      \step[fieldset=eprint, null]
    }  
  }
}

\DeclareMathOperator*{\argmin}{arg\,min}

\usepackage{graphicx}%
\usepackage{multirow}%
\usepackage{amsmath,amssymb,amsfonts}%
\usepackage{amsthm}%
\usepackage{mathrsfs}%
\usepackage[title]{appendix}%
\usepackage{xcolor}%
\usepackage{textcomp}%
\usepackage{manyfoot}%
\usepackage{booktabs}%
\usepackage{algorithm}%
\usepackage{algorithmicx}%
\usepackage{algpseudocode}%
\usepackage{listings}%
\usepackage{svg}

\newcommand{\consumerset}{\mathcal{B}}
\newcommand{\producerset}{\mathcal{S}}

\newcommand{\consumer}{b}
\newcommand{\producer}{s}

\newcommand{\coo}{CO\textsubscript{2}}

\begin{document}

\title[Article Title]{Quantum Optimization for the Future Energy Grid: Summary and Quantum Utility Prospects}


\author[2]{\fnm{Jonas} \sur{Blenninger}}
\author[2]{\fnm{David} \sur{Bucher}}
\author[1]{\fnm{Giorgio} \sur{Cortiana}}
\author[1]{\fnm{Kumar} \sur{Ghosh}}
\author[1]{\fnm{Naeimeh} \sur{Mohseni}}
\author[3]{\fnm{Jonas} \sur{Nüßlein}}
\author*[1]{\fnm{Corey} \sur{O'Meara}}\email{corey.o'meara@eon.com}
\author[2]{\fnm{Daniel} \sur{Porawski}}
\author[2]{\fnm{Benedikt} \sur{Wimmer}}


\affil*[1]{\orgdiv{E.ON Digital Technology GmbH, Hannover, Germany}}
\affil[2]{\orgname{Aqarios GmbH, Munich, Germany}}
\affil[3]{\orgname{Mobile and Distributed Systems Chair, Ludwig-Maximilians Universität, Munich, Germany}}


\abstract{In this project summary paper, we summarize the key results and use-cases explored in the German Federal Ministry of Education and Research (BMBF) funded project ``Q-GRID" which aims to assess potential quantum utility optimization applications in the electrical grid. The project focuses on two layers of optimization problems relevant to decentralized energy generation and transmission as well as novel energy transportation/exchange methods such as Peer-2-Peer energy trading and microgrid formation. For select energy grid optimization problems, we demonstrate exponential classical optimizer runtime scaling even for small problem instances, and present initial findings that variational quantum algorithms such as QAOA and hybrid quantum annealing solvers may provide more favourable runtime scaling to obtain similar solution quality. These initial results suggest that quantum computing may be a key enabling technology in the future energy transition insofar that they may be able to solve business problems which are already challenging at small problem instance sizes.
}




\maketitle

\section{Introduction}\label{sec1}
The "Q-GRID" project is a 3-year funded project by the German Federal Ministry of Education and Research (BMBF) awarded as part of the ``Anwendungsnetzwerk für das Quantencomputing" call in 2021. The project's goal is to investigate the basic applicability of quantum computing optimization techniques in the energy sector and, due to Noisy Intermediate Scale Quantum (NISQ) computers, identify the current major problem-specific challenges before practical development and deployment of fault-tolerant quantum computers. Current workflows in finding applied quantum computing use-cases typically involve small-scale toy models where researchers attempt to find the right balance between problem complexity and qubit usage due to limited resources. Formulating the problems in the correct manner often requires a trade-off in discretizing, simplifying and/or reducing the problem type or size all the while attempting to keep things realistic such that when larger hardware resources are available, the problem instances can still provide potential advantage and value. The central theme of the applied use cases of the project is to optimize various levels and aspects of a decentralized future energy grid system with particular attention being paid to problem scalability. The project focuses on two main application areas of the decentralized energy grid:

\begin{itemize}
\item \textbf{Load Scheduling using Dynamic Pricing}: How can utility companies encourage energy usage based on dynamic customer pricing to mitigate grid congestion while emphasising the use of energy during periods of high infeeds of renewable energy? 

\item \textbf{Virtual Community and Microgrid Creation for Sharing and Pooling Renewable Resources}: A microgrid is a subunit of the energy network grid which can become, to a large extent, self-sustained, provided when equipped with sufficient local renewable energy generation assets ($i.e.$ large or small scale PV installations). How can we optimally identify physical and virtual microgrids/communities based on current energy consumption and production? How can one share energy within an identified community to make optimal use of locally generated renewable energy?

\end{itemize}
In this project summary paper, we outline four proposed approaches for sub-problems related to the above overarching use-cases. We will describe them in terms of their background, mathematical formulation, and benchmarking results.

\section{Dynamic Price Incentivization for Renewable Energy Usage}
The increasing demand for energy resources and the growing use of renewable electricity have sparked a search for new ways to manage energy consumption and reduce grid congestion and carbon emissions. Demand Side Response (DSR)~\cite{DSR_00} is one such approach that actively adjusts energy usage based on grid conditions. Studies in the literature explore DSR's impact on smart grid technology \cite{DSR_01}, and load scheduling ~\cite{DSR_02}.
Energy price adjustment is one simple method to influence consumer behavior. Typically, this involves giving all customers a dynamic price simultaneously. However, because consumers have different usage patterns, alternative pricing policies may work better. Therefore, we aim to find personalized price patterns for each customer to optimize load distribution. We refer to this as Discount Scheduling Problem (DSP).
\subsection{Problem Formulation}
To formulate the problem, we discretize the optimization time horizon in $N_t$ steps and assign each customer $c = 1,\dotsc, N_c$ a discount (or penalty) $z_{c,t}$ at each timestep. Furthermore, we require forecasted consumption data $d_{c,t} \geq 0\,[\text{kWh}]$ for each customer and predicted grid \coo\ intensity $I_t\,[\text{g}/\text{kWh}]$ of the power generation in the considered region. The given discounts $z_{c,t}$ are defined as discrete discount steps. 
We employ \emph{integer encoding} by assigning each discount category $Z$ ($z_{c,t} \in Z$) to be discretized within a range $[z_\text{min}, z_\text{max}]$ into $N_k$ linearly spaced categories. We choose a symmetric interval $z_\text{max} = -z_\text{min} = z_m > 0$. 
Therefore, $Z = \{-z_m + i \Delta z\,|\, i = 0,\dotsc\,N_k-1\}$, with $\Delta z = \frac{2 z_m}{N_k-1}$. A $z_k < 0$ refers to a discount, while $z_k > 0$ is a penalty. This range can subsequently be expressed using $Q = \lfloor \log_2 N_k + 1\rfloor$ binary variables $x_{c,t,k}$ for each discount~$z_{c,t}$
\begin{gather}
    z_{c,t} = \Delta{z} \sum_{k=0}^{Q-1} w_k x_{c,t,k} - z_m, \\
    \text{with }w_k = \begin{cases}2^k &\text{if } k < Q - 1 \\ N_k - 2^{Q-1} + 1  &\text{else.} \end{cases}
\end{gather}
This encoding is space efficient, allowing for an exponential number of categories to be represented with a linearly growing number of qubits.


We choose the main goal of the optimization problem to be lowering the overall \coo\ emissions of all customers by shifting the load to times when the grid \coo\ intensity $I_t$ is small, and therefore, green electricity production is large. To this end, we introduce the average adjusted \coo\ grid intensity $\Delta I_t = I_t - \langle I_t \rangle_t$, which is negative whenever carbon emissions are low and positive otherwise. $\langle I_t \rangle_t$ is the average \coo\ intensity. The central assumption of the formulation is that a given discount (or penalty) influences the customer's behavior to alter their consumption proportionally. The new consumption then becomes
\begin{align}\label{eq:new_consumption}
    \tilde{d}_{c,t} = (1 - \chi_c z_{c,t}) d_{c,t},
\end{align}
when given a discount $z_{c,t}$, where $\chi_c$ is the (negative) price elasticity of demand of customer $c$. i.e., the higher $\chi_c$ is, the more customer $c$ responds to price incentives (lower its demand if price increases and vice versa). Finally, we can formulate the cost function for our optimization $\argmin_{z} R(z)$ as follows
\begin{align}\label{eq:co2_reduction}
    R(z) = \sum_{c,t} \Delta I_t [1 - \chi_c z_{c,t}]d_{c,t},
\end{align}
where $z(x) \in Z^{N_c \times N_t}$ is the discount matrix that can be encoded through binary variables $x\in \mathbb{B}^{N_c \times N_t \times Q}$. This objective function is subject to several constraints such as not wanting customers to change their total consumed energy over the optimization horizon, and, enforcing the momentary change in consumption of all customers to be bounded, to preserve system stability. To tackle large problem instances, a use-case specific problem decomposition technique has been developed. We refer the reader to \cite{bucher2023dynamic} for further details on the mathematical structure of the constraints and of the decomposition method.

\subsection{Results}

As a state-of-the-art purely classical baseline, we used \texttt{Gurobi}\footnote{All experiments with Gurobi were conducted on an M1 MacBook Pro (2020) with Gurobi Version 9.0}~\cite{gurobi}. This was compared to D-Wave’s \texttt{LeapHybridCQM} solver~\cite{mcgeoch} (subsequently referred to as \texttt{Leap}), which is a quantum-classical hybrid algorithm that uses classical algorithms to optimize the problem whilst using quantum (annealing) computers to solve suitable sub-tasks. This has the benefit of solving larger problems than possible directly on current quantum hardware while also supporting more sophisticated optimization models that include hard constraints. \texttt{Leap} is accessed through D-Wave's Cloud service. These two out-of-the-box solvers are compared against our own problem-specific decomposition routine \cite{bucher2023dynamic}, subsequently called \texttt{Decomp-Gurobi}, \texttt{Decomp-Leap} or \texttt{Decomp-QPU}, depending on the method considered for solving the chunk problems. \texttt{QPU} refers to direct access to the D-Wave's Quantum Annealing processor Advantage 4.1~\cite{mcgeoch}. 
Whenever a decomposition solver is followed by an integer, it refers to the split size $m$ indicating a subroutine solver is ran on $m$ number of customers.

Because we are considering an optimization task with multiple goals involved, it is not sufficient to consider only the objective value of our model as a performance metric. Instead, we simultaneously investigate multiple metrics.
\emph{\coo\ reduction}: First and foremost, the \coo\ reduction is the central goal of the DSP, hence it is also the main metric that is investigated. \emph{Energy}: The energy, or objective, of the optimization problem consists of the rescaled \coo\ reduction with the penalties added.  
\emph{Average relative cost savings}: How evenly distributed are the discounts across all customers?

If we want to solve the DSP for a given data frame, consisting of the consumption of $N_c$ customers at $N_t$ time steps, we still need to fix a set of open variables and parameters. In a real-world scenario, the customer price elasticity on demand $\chi_c$ could be measured from the individual customer's behavior. However, as it only acts as a proportionality constant, we turn their effect off and set them all to one. Next, we use five discount categories, with a 50\% discount maximally. That, in turn refers to the following valid discounts $z_{c,t} \in \set{-50\%, -25\%, 0\%, 25\%, 50\%}$. As a consequence, a discount of, e.g., 50\% would result in an increase in the customer's consumption by 50\%. Additionally, the precise parameter settings for each of the solvers can be found in \cite{bucher2023dynamic}.

\begin{figure}
    \centering
    \includegraphics[width=0.45\textwidth]{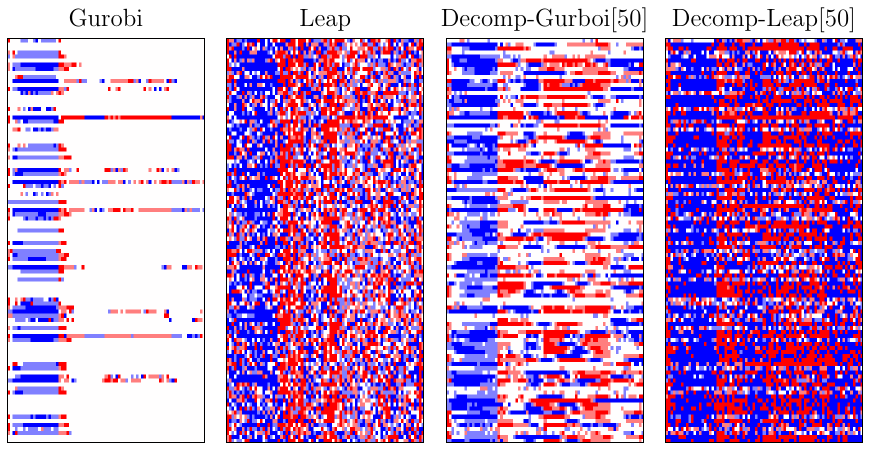}
    \caption{The discount matrices $z_{c,t}$ found by the investigated solvers for $N_c = 100$. Blue indicates a discount and red corresponds to a penalty. White means no discount given at all. Despite their effects on the overall consumption (see Fig.~\ref{fig:ex100-consumption}) being the same, the discount matrices are significantly different from one another. It is apparent that the \texttt{Gurobi} solver provides a more greedy approach to discount allocation than \texttt{Leap}, thereby indicating a larger impact of the regularization. Nevertheless, a similar pattern is observable in the last three solutions.} 
    \label{fig:ex100-discount-matrix}
\end{figure}

\begin{figure}
    \centering
    \includegraphics[width=0.4\textwidth]{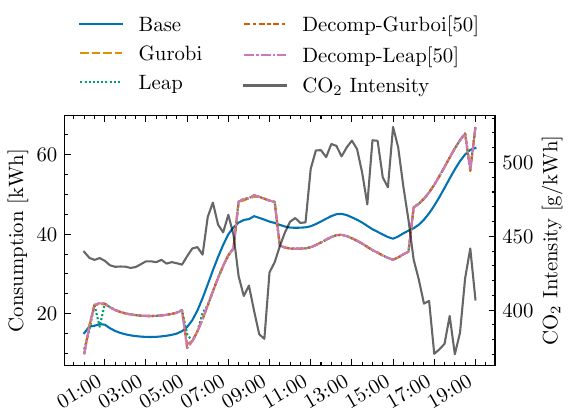}
    \caption{The effect of the DSP solution for problem size $N_c = 100$. The plot shows the aggregated consumption with and without (Base) discounts in place, as well as the grid \coo\ intensity. The solutions of all solvers produce the same (effectively) consumption change. Times with high \coo\ emissions produce an effective decrease in consumption and vice versa, as expected. }
    \label{fig:ex100-consumption}
\end{figure}

Let us first examine the optimization result of the different solvers in detail for a 100-customer example before focusing on the previously discussed metrics.
We analyze the solutions of four solvers, \texttt{Gurobi}, \texttt{Leap}, and two $m = 50$ decomposition methods with the same solvers as the sub-routine. 
The results for the discount matrices $z_{c,t}$ can be seen in Fig.~\ref{fig:ex100-discount-matrix}, while their overall effect on the consumption is displayed in Fig.~\ref{fig:ex100-consumption}. 
Although the particular solutions differ significantly from one another, the effective overall energy result stays similar (essentially the same), regarding the \coo\ reduction. The difference between \texttt{Gurobi} and the other solvers is notable in terms of the allocated individual customer dynamic tarifs.

\begin{figure}
    \centering
    \includegraphics[width=0.35\textwidth]{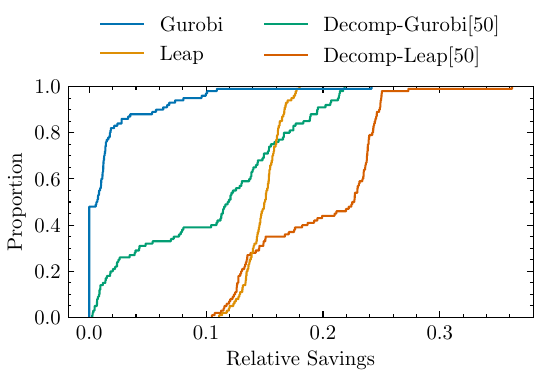}
    \caption{A cumulative distribution plot of the relative savings of the customers. \texttt{Gurobi} provides a more greedy discount allocation approach by providing discounts to relatively few customers. On the other hand, \texttt{Leap} distributes similar discounts to all customers. Remember: We do not optimize for this metric. This is just an observation of the different strategies and can be interpreted as a measure of the fairness of the optimization algorithms with respect to different customers.}
    \label{fig:ex100-fairness}
\end{figure}

\begin{figure}
    \centering
    \includegraphics[width=0.45\textwidth]{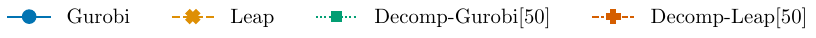}\\
    \includegraphics[width=0.24\textwidth,trim=0 15 0 0,clip]{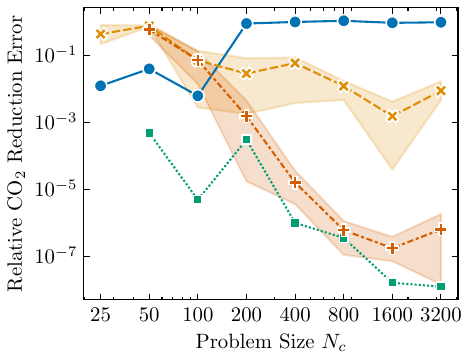}%
    \includegraphics[width=0.24\textwidth,trim=0 15 0 0,clip]{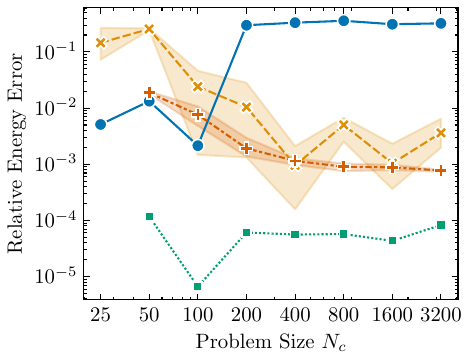}\\
    \caption{The investigated metrics for different problem sizes and different solvers. The runtime of all solvers has been set to be equal for a certain problem size, but grows with $N_c$. The top row shows the global metrics, which tell the most about how the solver performed. The relative energy error is the central objective that we try to minimize, while the energy error gives an overview of the performance with regard to all optimization targets.}
    \label{fig:eval-full}
\end{figure}

The results for two key metrics previously discussed are visualized in Fig.~\ref{fig:eval-full}. Each plot shows a singular metric against the problem size for the considered solvers. Focusing on the relative \coo\ reduction, it is evident that a crossover in performance between \texttt{Gurobi} and \texttt{Leap} happens between 100 and 200 customers. After that size, \texttt{Gurobi} cannot find converged results in the given time limit. Although not a directly fair comparison since \texttt{Gurobi} is run on a local machine whilst the \texttt{Leap} hybrid solver is run on a proprietary D-Wave cloud architecture. Nevertheless, the classical decomposition routine greatly outperforms the general-purpose solvers, and in the higher customer size problem sizes, the decomposition method with Leap as subsolver remains competitive. Apart from the global optimization action, we are also interested in how the optimization performs per customer. Figure \ref{fig:ex100-fairness} demonstrates that there seems to be a more even application of discounts to the customer set by the \texttt{Leap} solver. For a more in-depth analysis please see \cite{bucher2023dynamic}.

\section{Self-sufficient Energy Community Detection}

In this use-case, we focus on identifying self-sufficient microgrids within an ever-changing (based on generation and consumption) distribution grid consisting of multiple consumers and producers of electricity. Some of the actors in the grid are consumers and producers combined, so-called \emph{prosumers}, via self-installed photovoltaic panels, home batteries, or Vehicle-to-Grid technologies.

To this end, we formulate the problem as a Quadratic Unconstrained Binary Optimization (QUBO) task to run the problem on quantum computing hardware. We also develop two heuristics---one classical and one hybrid with QUBO sub-problems---to tackle the problem more efficiently. Our approach is built on Community Detection, a well-known NP-hard problem with many use cases~\cite{brandes2006}. The general Community Detection problem has been investigated via the use of Quantum Annealing (QA) already in the literature~\cite{fernandez-campoamor2021b,reittu2019, shaydulin2019a, negre2020a,gemeinhardt2021a, stein2023}, however here we focus on augmenting the cost function via additional constraints, which as we will show, make the problem substantially harder classically at smaller instance sizes - a prerequisite for well formulated near-term quantum utility applications \cite{abbas2023quantum}.

\subsection{Problem Formulation}

Community Detection relies on finding a community assignment that maximizes a quantity called \emph{Modularity}. A community structure of a graph with $N$ nodes and (max) $K$ communities can be expressed through $N \times K$ binary variables, where $x_{i,\ell} = 1$ if node $i$ belongs to the community $\ell$. The binary variables must obey so-called one-hot encoding constraints, meaning that only a single variable can be non-zero for one node. Community detection through modularity maximization is defined as follows~\cite{brandes2008}
\begin{gather}
    \max \frac{1}{2m}\sum_{i,j} \left(A_{i,j} - \frac{k_i k_j}{2m} \right) \sum_{\ell} x_{i,\ell} x_{j,\ell},
\label{eq:cd-qubo}
\end{gather}
where $A_{i,j}$ is the adjacency matrix of the network, $k_i = \sum_j A_{i,j}$, and $m = \frac{1}{2} \sum_i k_i$. A peculiarity in our approach is that we dynamically compute the weights of the adjacency matrix based on the current load and production in the grid. Each consumer or prosumer has a node power value $p_i$, where $p_i < 0$ refers to production and $p_i > 0$ to consumption. With this data, we can compute the power flowing through the grid and use the power that flows through a specific line as the weight in $A_{i,j}$. Doing so will collect customers, where the physical power flows, into communities.

The self-reliance condition can be additionally accounted for by minimizing the penalty term
\begin{align}
    \min_x \frac{\lambda}{\mathcal{P}} \sum_\ell \left( \sum_i p_i x_{i,\ell} \right)^2,
\end{align}
where $\mathcal{P}$ is a normalization constant and $\lambda$ is a penalty factor that steers how much self-reliance is respected when combined with Eqn.~\eqref{eq:cd-qubo}.

Inspired by the GCS-Q algorithm which was developed for coalitional game structure generation~\cite{venkatesh2023gcs}, we apply a divisive algorithm (a greedy heuristic) that tries to find the best community structure with $K = 2$, i.e., splitting the network in two. This is recursively repeated in the two sub-nets until the solution quality cannot be further improved. By optimizing the problem only for $K = 2$, we can encode the first group as $x_i = 0$ and the second as $x_i = 1$, circumventing the need for a one-hot-constraint. Therefore, the largest QUBO problem consists of $N$ binary variables and reduces in size throughout the algorithm. Furthermore, as we will also see in the benchmarking section, the sub-problems are generally more manageable for the solvers since no constraints need to be fulfilled.

\subsection{Results}


\begin{figure}
    \centering
    \includegraphics[width=0.4\textwidth]{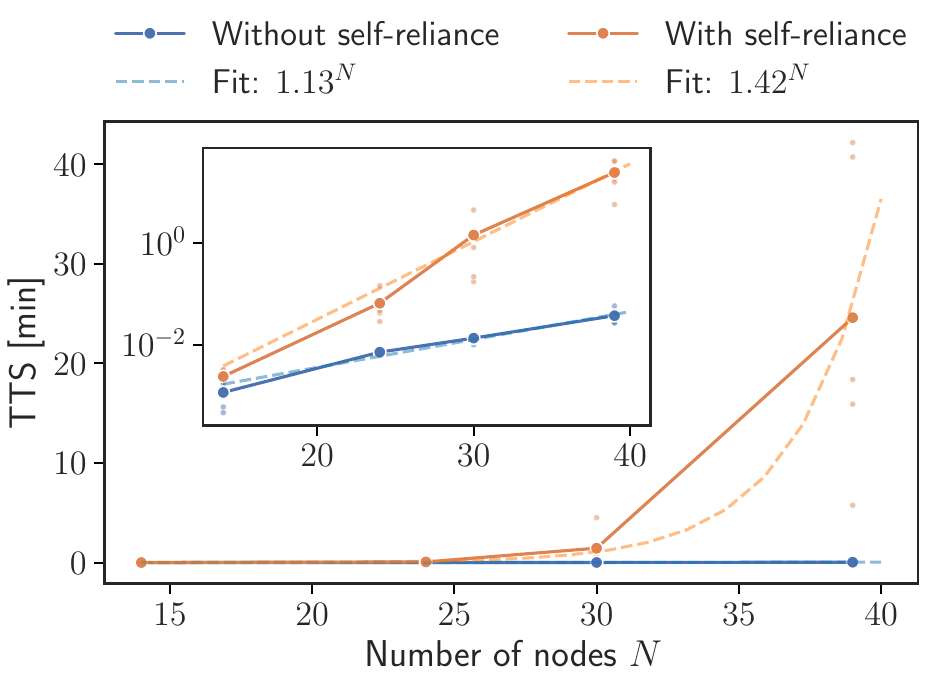}
    \caption{The time to solution of Gurobi on the Single-Problem showcases the exponential difficulty of the problem, especially when self-reliance is considered.}
    \label{fig:bench-and-tts}
\end{figure}

\begin{figure*}
    \centering
    \begin{subfigure}{0.3\textwidth}
    \includegraphics[width=\textwidth]{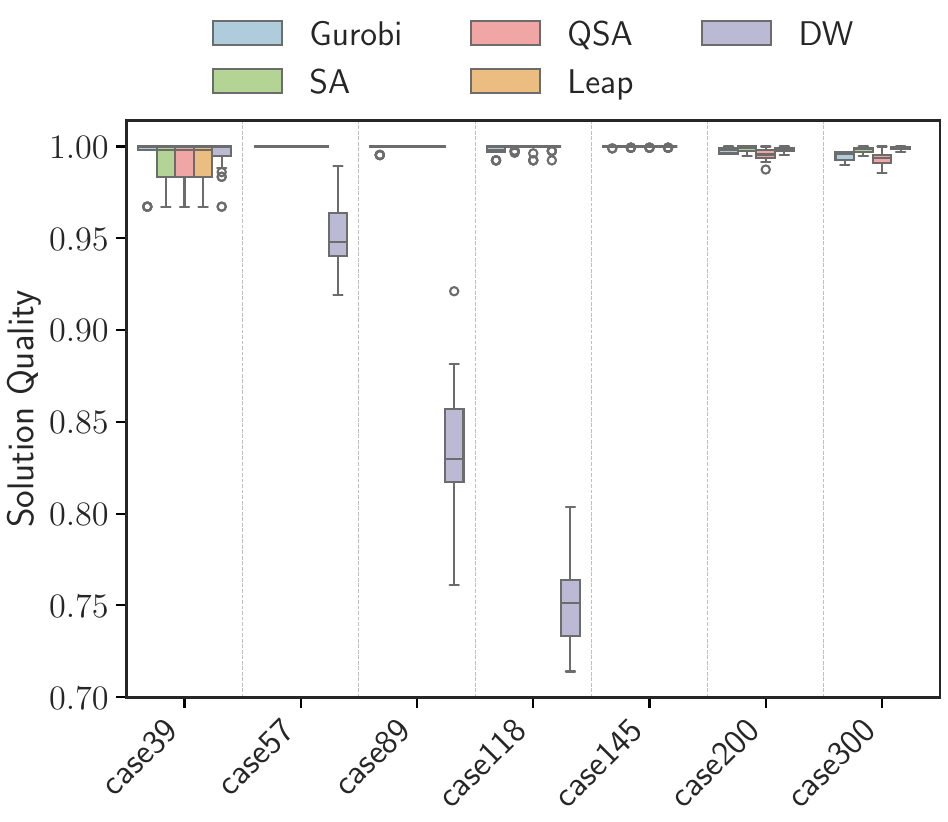}
    \subcaption{Divisive Approach}
    \label{fig:benchmarking-gcs}
    \end{subfigure}
    \begin{subfigure}{0.32\textwidth}
    \includegraphics[width=\textwidth]{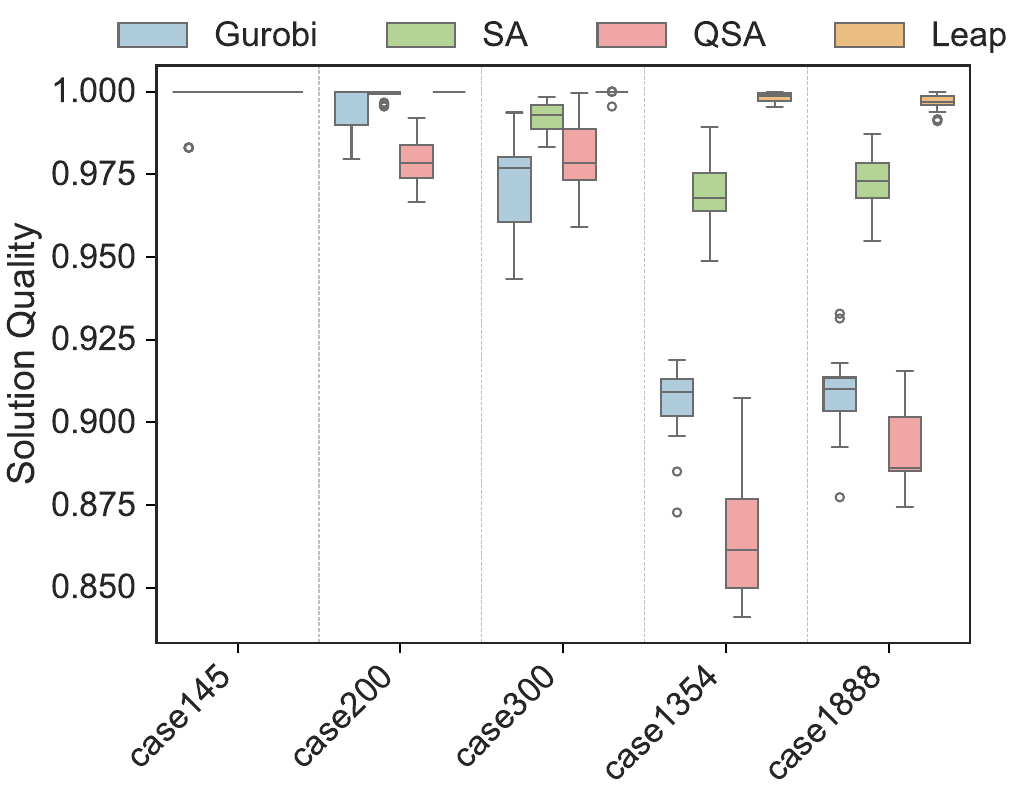}
    \subcaption{First Split of Divisive}
    \label{fig:benchmarking-fs}
    \end{subfigure}
    \begin{subfigure}{0.35\textwidth}
    \includegraphics[width=\textwidth]{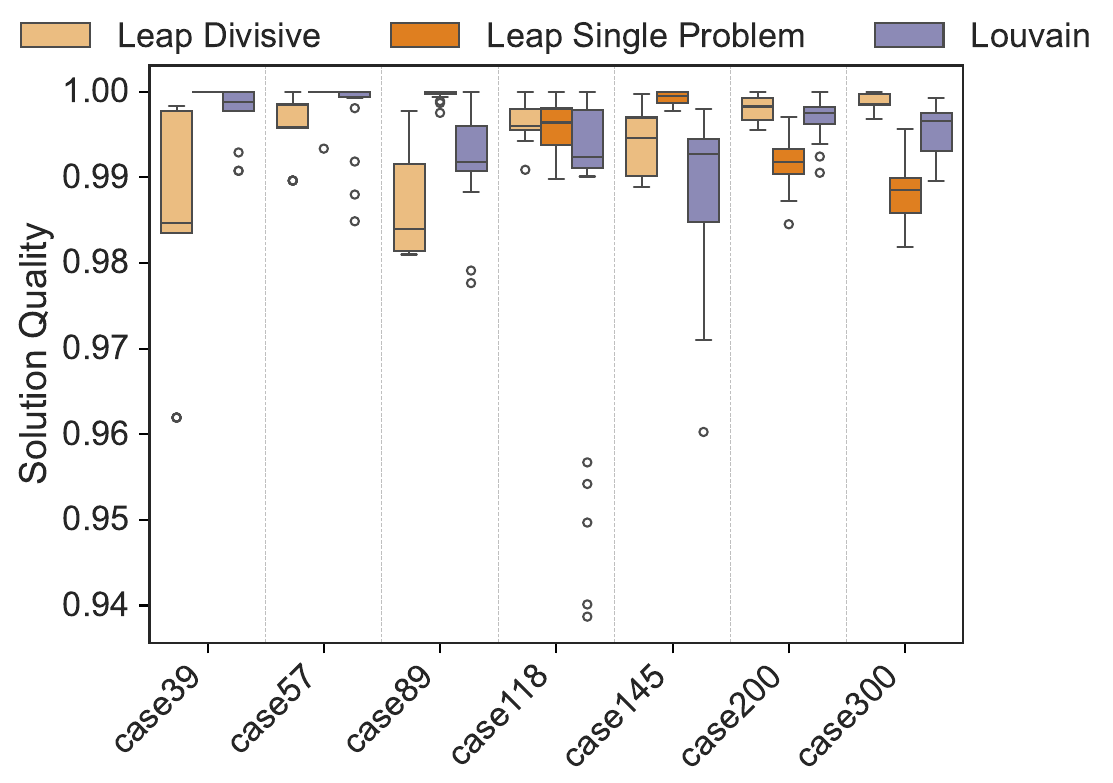}
    \subcaption{Comparison of Leap and Louvain}
    \label{fig:benchmarking-leap}
    \end{subfigure}
    \caption{Results of the SRCD Benchmarking, the divisive approach in (a), the first sub-problem of the divisive approach in (b), and a focused comparison of Leap with both approaches and Louvain.}
    \label{fig:benchmarking}
\end{figure*}
Figure \ref{fig:bench-and-tts} demonstrates the exponential runtime nature of the problem. Here, Gurobi is run until convergence for self-reliance turned on or off, clearly showing that the problem becomes harder with self-reliance. The benchmarking grid instances comprise of IEEE power system test cases provided by \texttt{pandapower}~\cite{pandapower} and have been adapted to behave more like residential power grids. The consumption and production data is sampled randomly from two shifted normal distributions (one for production and one for consumption). 50\% of the actors are producers and the other 50\% are consumers. For each test case, we sample five data instances. The self-reliance penalty $\lambda$ is set to 0.1 due to preliminary trials which demonstrated the best trade-off with this setting. Furthermore, the number of communities $K$ is estimated using Louvain's algorithm, by running the heuristic and then choosing the median number of communities which were found.
The considered solvers in the benchmark comparison are; \emph{Gurobi}~\cite{gurobi}: a classical branch-and-bound MIP solver, \emph{Leap}~\cite{mcgeoch}: a cloud-based hybrid quantum optimizer by D-Wave (CQM for single-problem and BQM for divisive), \emph{SA}~\cite{sa} and \emph{QSA}: QUBO heuristics Simulated Annealing and QBsolv with Simulated Annealing, \emph{Louvain}: Adapted classical Community Detection heuristic, and \emph{Divisive}: With Gurobi, Leap, SA, QSA, and D-Wave's QPU as QUBO solvers.
Each solver is run five times for a single instance, resulting in 25 runs for one case. In the single-problem runs, the timeout is fixed to 6\,s. In the Divisive runs, each QUBO solution is given 3\,s timeout (slightly increasing for cases with more than 1000 nodes, as the minimum runtime for the Leap solver rises). We use a relative solution quality that normalizes the energy by the best-found energy for one instance. 

Figures \ref{fig:benchmarking-gcs} and \ref{fig:benchmarking-fs} demonstrate the performance of the divisive approach together with the performance on the first sub-problem (on more substantial problem instances). The divisive approach delivers good results over almost all solvers, except for QA, which loses solution quality immediately. 
The first split results on the larger instances give a more detailed picture of the trend of the remaining solvers, rendering Leap as the best performer, followed by SA, Gurobi, and QSA. Lastly, Fig.~\ref{fig:benchmarking-leap} shows the performance comparison of Leap (both approaches) and Louvain. Except for the smaller cases, one of the Leap approaches always slightly outperforms Louvain. 

\section{Prosumer Coalition Formation\label{sec:introduction}}
Instead of explicitly determining microgrid communities which have balanced energy consumption and generation, we can instead approach the community detection problem from a cooperative game theoretic approach. In the context of energy community formation, cooperative game theory provides a framework for understanding how prosumers can form coalitions to optimize their collective benefits, for example to pool resources together to sell bulk energy as a ``team" \cite{bandeiras2023application, moafi2023optimal, han2018incentivizing}. Within this framework, a coalition structure pertains to the arrangement of agents into groups, denoted as ($C$). A coalition can be characterized by a characteristic function that measures the cooperative advantage, typically denoted as $\nu (C)$, which assigns values to different coalitions \cite{bachrach2013optimal, deng1994complexity}. 

The task of creating coalition structures is highly challenging, especially given the presence of $n$ agents, leading to a staggering $O(n^n)$ possible partitions. Currently, the only algorithm capable of finding an optimal solution within a reasonable computational complexity is the Dynamic Programming algorithm, achieving this in $O(3^n)$ time \cite{rahwan2008improved}. The problem can be transformed into an approximately equivalent game represented using the Induced Subgraph Game (ISG). In this representation, nodes correspond to agents, and coalition values are reflected in pairwise interactions within a weighted graph, such that $v\left(C_l\right)=\sum_{(i, j) \in C_l} w_{i j}$ \cite{deng1994complexity, bachrach2013optimal} with $w_{i j}$ as weights of the graph. The transformation of to ISG allows the problem to be recast as quadratic unconstrained binary optimization, enabling more computationally efficient approaches. However, it is important to note that the problem remains NP-complete \cite{bachrach2013optimal}. In this use-case, we employ approximate algorithms introduced in \cite{venkatesh2023quacs, mysore2022gcs}, and benchmark against exact classical solvers with the Quantum Approximate Optimization Algorithm (QAOA).  


\subsection{Problem Formulation}

Consider a distribution network with $N$ prosumers, each possessing inflexible loads, flexible loads, generation, and/or storage. Retail contracts typically involve individual metering where prosumers aim to minimize costs, balancing energy needs using flexible resources, subject to constraints. Individual metering encourages prosumers to optimize independently, reducing upstream power flows. However, it lacks coordination between prosumer groups. An alternative is forming net-metered energy communities, where collective net demand is metered. This promotes community supply-demand balancing, modeled by a collective energy management problem \cite{8417894}.


In this optimization framework, prosumers are partitioned into coalitions, aiming to maximize the sum of values associated with each coalition. The decision variable $x_l$ represents the inclusion of coalition $C_l$ in the selected partition, subject to constraints. The objective function considers the value of each coalition in terms of reduced network management costs. The Distribution System Operator (DSO) decision problem is then to partition the prosumers into a set of net-metered coalitions with the goal of minimizing network management costs. This is formulated as a Coalition Structure Generation (CSG) problem, which can be solved using Binary Integer Linear Programming (BILP)~\cite{venkatesh2023gcs},
\begin{equation}
\begin{aligned}
\max _{x_l, l \in\left[2^n-1\right]} & \sum_{l \in\left[2^N-1\right]} v\left(C_l\right) x_l, \\
\text { s.t. } & \sum_{j \in\left[2^N-1\right]} S_{i l} x_l=1, \forall i \in[N], \\
& x_l \in\{0,1\}, \forall l \in\left[2^N-1\right] .
\end{aligned}
\end{equation}

Note that there are $2^N-1$ potential sub-coalitions of prosumers. $S_{i l}=1$ if prosumer $i \in[N]={1,2,...,N}$ belongs to coalition $C_l$ (i.e. $i \in C_l$ ), and $S_{i l}=0$ otherwise. Binary decision variable $x_l=1$ indicates that $C_l$ is part of the selected partition, and the value function $v\left(C_l\right)$ gives the value of the coalition.
This formulation is computationally intensive since the number of binary decision variables depends on the number of sub-coalitions, which increases exponentially with the number of prosumers (i.e. $2^N-1$ ). This CSG formulation can be transformed into ISGs, where the value of a coalition can be expressed as the sum of pair-wise ``joint utilities" between prosumers $w_{i j} \in \mathbb{R}, i, j \in[N]$, i.e.$
v^{i s g}\left(C_l\right)=\sum_{(i, j) \in C_l} w_{i j}$ with lower computational requirements.

The value function nonlinearities in $v(C_l)$ mean this framework is not directly applicable to our CSG. However, we can make use of an ISG based on pairwise joint utilities $\tilde{w}_{i j}$ which approximate our GSCP. The following quadratic programming problem can be used to find pairwise weights that minimize the mean squared error between the value functions of the GSCP and an approximate ISG,
\begin{equation}
\min _{\tilde{w}_{i j}, i, j \in[N]} \sum_{l \in\left[2^N-1\right]}\left(v\left(C_l\right)-\sum_{(i, j) \in C_l} \tilde{w}_{i j}\right)^2 .   
\end{equation}

Starting with the grand coalition, an approximate solution to the ISG can be found by iteratively splitting coalitions in two, until no value-increasing bipartitions are available \cite{venkatesh2023quacs, venkatesh2023gcs}. For a given iteration, let $C \in[N]$, be one of the optimal bipartitions that was found. The optimal bipartitions of $C$ at the next iteration can be found using the following QUBO with $N$ binary decision variables,
\begin{equation}
\begin{aligned}
\min _{\tilde{x}_i, i \in C} & \sum_{i \in[N]} \sum_{j \in[N] \backslash i} \tilde{w}_{i j} x_i\left(1-x_j\right) . \\
\text { s.t. } & x_i \in\{0,1\}, \forall i \in[N] .\label{eq1}
\end{aligned}   
\end{equation}

\subsection{Results}
\begin{figure}[t]
\centering
\includegraphics[width=1\linewidth]{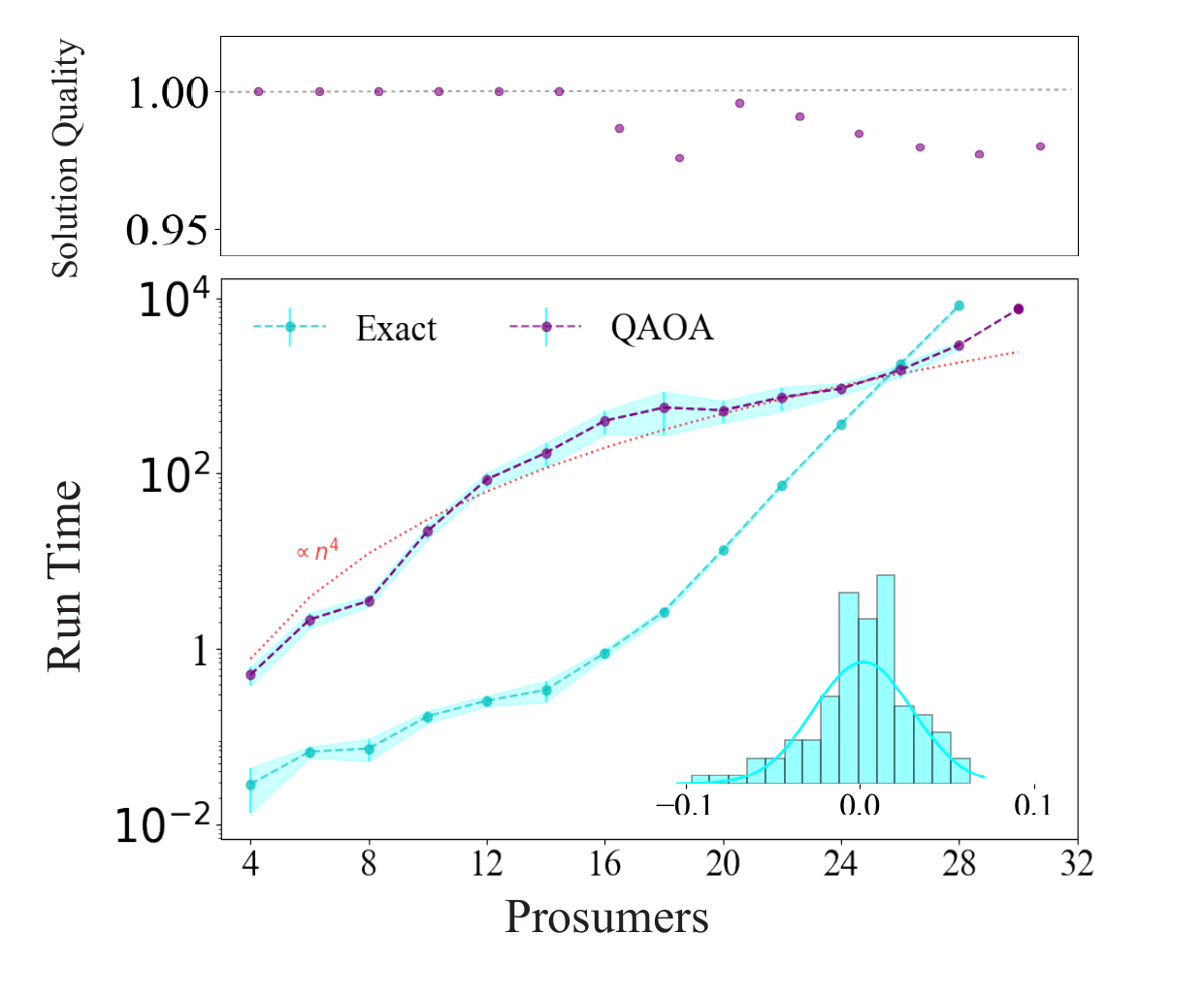} 
\caption {Benchmarking an exact classical solver against QAOA. Results for each number of prosumers showcase an average across 20 problem instances.  The light blue histogram  in the lower right shows the weight distribution after
transforming CSG to ISG.
\label{fig2}}
\end{figure}
In this section, we elaborate on our benchmark between an exact classical solver and  1-round QAOA to find optimal bipartitions through iterative minimization of Eq. \eqref{eq1}. Our emphasis lies on evaluating solution quality and run time as performance metrics. The solution quality is determined by calculating the ratio between the value of the optimized coalition obtained by 1-round QAOA and that achieved by the exact classical solver. 

Before delving into performance metrics, we examine the weight distribution resulting from the CSG to ISG mapping, as illustrated in Fig. \ref{fig2} (The lower right histogram). The weight distribution showcases an average across 20 realizations for the case of 18 prosumers.  Note that distinct realizations correspond to varied initial power for prosumers at the beginning. Our observations reveal that the graphs generated from transferring CSG to ISG are nearly fully connected across all system sizes. Furthermore, the weight distribution exhibits a comprehensive range, incorporating both positive and negative values, thereby forming a Gaussian distribution centered close to zero. 

The performance results are shown in Fig. \ref{fig2} where the number of shots across all prosumer counts is fixed to $2^{12}$. The exact classical solver, due to its computational complexity, experiences a noticeable increase in runtime that becomes impractical beyond 30 prosumers. In contrast, while 1-round QAOA does not guarantee an exact solution, it demonstrates a more favorable scaling in terms of running time, showcasing its potential efficiency in handling larger prosumer counts in comparison with the exact solver. These results demonstrate a potential competitive advantage for quantum-based solvers compared to exact solvers in terms of problem size scaling to runtime while maintaining a high solution quality.


\section{Peer-2-Peer Energy Trading}
Once self-sufficient communities or coalitions have been identified across the electrical grid, the act of trading or scheduling energy exchange between community members needs to be considered. In these microgrids, the role of the producer and consumer is not static. Participants might produce excess energy that can be sold to other participants or lack energy requirements and need to buy the excess energy from other participants. In previous work~\cite{10129617}, the authors propose to match the prosumers based on their ask and bid price to maximize the social welfare of the participants. The matching is run by an auctioneer optimizing for the maximum overall traded value while ensuring that no consumer buys more energy than requested and no producer sells more energy than generated. In addition, the auctioneer ensures that buyers do not undercut the price of the matched seller. A significant drawback of this matching approach is that the physical power flow in the grid is not respected. The power of the matched participants by the auctioneer might not flow from the producers to their matched consumers but from and to any of the other participants.
This mismatch of the logical power flow to the physical power flow can result in an unfair disadvantage for producers closer to their potential consumers being undercut by producers further away in the grid. 

\subsection{Problem Formulation}
We propose a problem formulation matching the grid participants in a Peer-2-Peer auction so that the resulting logical power flow is as close to the actual physical power flow in the grid as possible.
Let us describe the grid as a directed weighted graph $G(V, E, e)$, where $V = \producerset \cup \consumerset$ are the vertices (or nodes or prosumers) and $E \subseteq V \times V$ are the directed edges or power lines. The edge weight $e_{u, v}$, with $(u, v) \in E$, indicates the physical power flow from prosumer $u$ to prosumer $v$ over the particular line in kW. We are interested in matching prosumers so that the logical power flow between the matched prosumers closely resembles the physical power flow in the grid.

For a producer $s \in \producerset$ and a consumer $b \in \consumerset$, we must consider all possible simple paths (paths without loops, i.e., paths without repeating vertices) from $s$ to $b$. 
The path sets $T_{\producer, \consumer} = \{t^1_{\producer, \consumer}, \dots, t^{n_{\producer, \consumer}}_{\producer, \consumer}\}$ include all possible $n_{\producer, \consumer}$ paths. 
Each path consists of the edges that lead from the producer to the consumer $t_{\producer, \consumer}^i = \{(\producer, v_1), (u_2, v_2), \dots, (u_{n_{\producer, \consumer}^{i}}, \consumer) \}$. 
The paths are calculated on the undirected graph of the grid, i.e., without the direction information of the physical power flow in the grid.
This allows us to model the logical power flow in the grid more accurately since power flowing in the same direction as the physical power flow might get canceled out or be reduced by power flowing in the opposite direction.

In addition to the direction and the physical power flow on a power line, each power line within the network has an associated resistance $\tilde{r}_{u,v}$, such that we may formulate a total resistance of a path as
\begin{align}
    r(t_{\producer, \consumer}^i) = \sum_{(u,v) \in t^i_{\producer,\consumer}} \tilde{r}_{u,v}.
\end{align}
The total resistance of all parallel lines is given by a reciprocal addition, i.e.,
\begin{align}
    \frac{1}{R_{\producer, \consumer}} = \sum_{t \in T_{\producer, \consumer}} \frac{1}{r_{\producer,\consumer}(t)}.
\end{align}
If we have the power $q_{\producer, \consumer} = \min(Q_\producer, Q_\consumer)$ flowing from the producer to the seller, we can compute the share of power flow that runs through each line as
\begin{align}
    w_{\producer,\consumer}(t) = \frac{R_{\producer,\consumer}}{r_{\producer,\consumer}(t)} \in [0, 1],
\end{align}
with $\sum_{t \in T_{\producer, \consumer}} w_{\producer,\consumer}(t) = 1$.
In addition to the share of power flowing through each line, we also need to calculate if this share flows in the same direction as the physical power flow or in the opposite direction. So for an edge $(u, v) \in E$, we can calculate the direction of the power flow of this particular edge in a path $t \in T_{s, b}$ as
\begin{align}
    d_{u,v}(t) = \begin{cases}
        +1 & \text{if }(u, v) \in t \\
        -1 & \text{if }(v, u) \in t \\
         0 & \text{otherwise},
    \end{cases}
\end{align}
where $+1$ indicates the logical power flow of the edge $(u, v)$ in the path $t$ flows in the same direction as the physical power flow, $-1$ indicates the logical power flow flows in the opposing direction as the physical power flow and $0$ otherwise, i.e., if the edge $(u, v)$ is not an element of the path $t$ between $s$ and $b$.
Thus, the logical power flow contribution on a line $(u,v) \in E$ of a trade between $\producer$ and $\consumer$ is given by
\begin{align}
    q_{\producer,\consumer}^{u,v} = q_{\producer,\consumer} \sum_{t \in T_{a,b}} w_{s,b}(t) d_{u,v}(t).
\end{align}
And, with the decision variables, if a trade happens $x_{\producer,\consumer} \in \{0,1\}$, we can compute the aggregated logical power flow over the line $(u,v) \in E$ as follows
\begin{align}
    q^{u,v}(x) = \sum_{\consumer \in \consumerset, \producer \in \producerset} q_{\consumer,\producer}^{u,v} x_{\producer, \consumer}.
\end{align}
Finally, we can formulate the objective function as the minimization of the mismatch between physical and logical power flow:
\begin{align}
    \min_{x_{\consumer,\producer}}  \sum_{(u,v) \in E} \left(e_{u,v} - q^{u,v}(x)\right)^2.
\end{align}

\subsection{Results}
We evaluate our problem formulation on several different benchmark grids, referred to as \textsf{cases}, of varying size and structure from the \texttt{pandapower} library~\cite{pandapower}. 
The largest grid evaluated consisting of $39$ prosumers is visualized in Fig.~\ref{fig:matches-in-grid}. 
We optimize for the optimal matching using three different solvers: \textit{Gurobi}~\cite{gurobi}, Simulated Annealing (\textit{SA})~\cite{sa}, and Tabu Search with Simulated Annealing on sub-problems (\textit{QSA})~\cite{hybrid-overview, sas}. The SA and QSA solvers are executed with the number of reads set to $1000$.
We run the optimization $10$ times for each case and solver with a maximum run time of $30$ seconds. 

\begin{figure*}
\centering
{\includegraphics[width=0.8\linewidth]{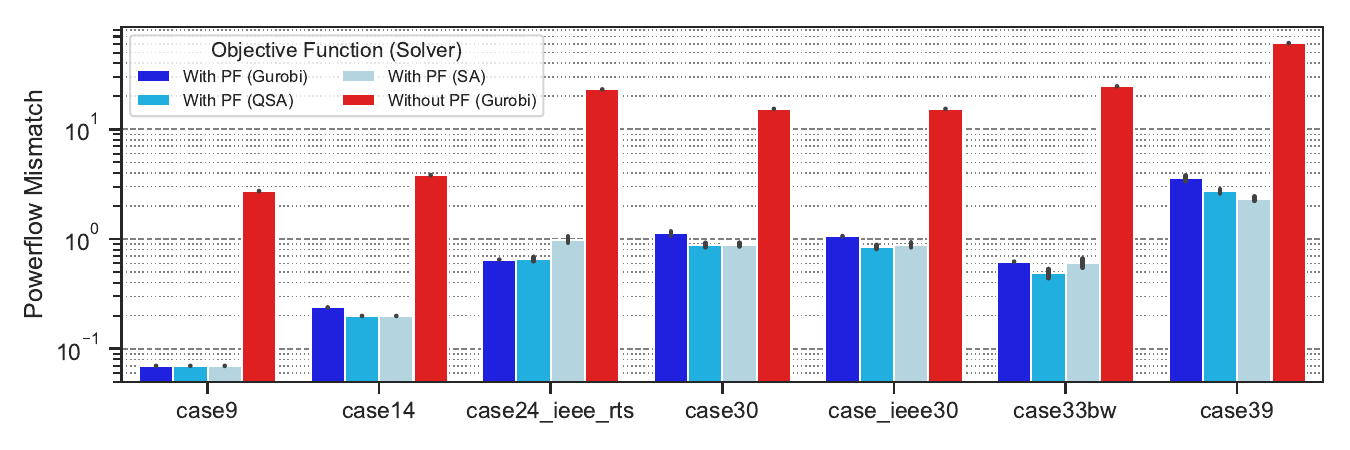}}
\caption{Comparison of the power flow mismatch of the PF formulation (blue) for three different solvers and the optimal solution of the non-PF formulation (red) over the cases. The closer the objective value is to zero, the better the results.}\label{fig:aq-obj-value-comparison}
\end{figure*}

\begin{figure}
\centering
{\includegraphics[width=0.8875\linewidth]{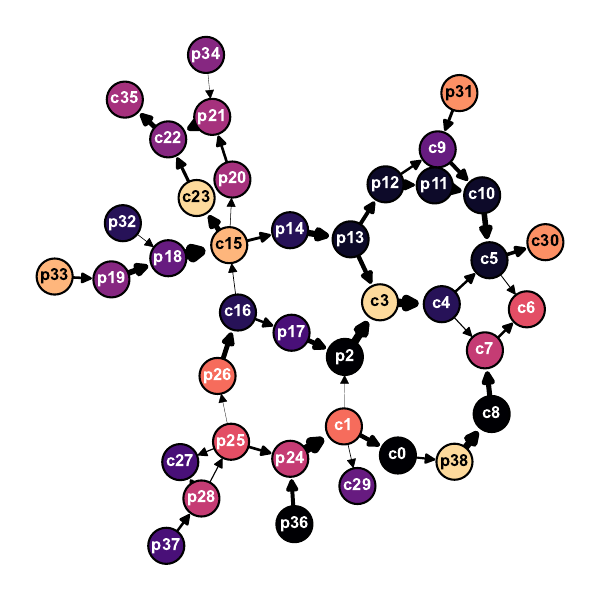}}
\caption{The physical power flow in the grid of \textsf{case39} with nodes colored according to the matched participants obtained. }
\label{fig:matches-in-grid}
\end{figure} 

Inspecting the results given in Fig. \ref{fig:aq-obj-value-comparison}, we can see that the differences in the solution quality for the examined solvers are marginal, i.e., we no longer obtain significantly better objective values with Gurobi as we do with the quantum-inspired solvers SA and QSA. As expected, compared to the optimal solution of the non-PF formulation given in~\cite{10129617}, the PF formulation exhibits a far smaller powerflow mismatch. The PF formulation is harder to solve classically, and the quantum-inspired heuristics deliver satisfactory results. These results demonstrate that matching a grid's participants based on the physical power flow in the grid can significantly increase the fairness of the matched prosumers regarding how the power flows in the grid. However, in our current formulation, the energy price is not respected, possibly resulting in consumers overpaying for the energy they consume. A solution to this could be to extend our proposed formulation with a price matching similar to the one suggested in~\cite{10129617}. 

\section{Conclusion}
In this project summary paper, we have outlined 4 industry-related use-cases for applied quantum optimization to the energy sector. In each of the use-cases, we have benchmarked the performance of the quantum algorithms versus classical counterparts. We have demonstrated that quantum optimization approaches may provide some advantages over traditional classical optimization approaches in terms of application-specific benchmarks or hybrid algorithm solution quality, however further investigation is needed to explore any evidence of potential quantum utility in the energy sector.
\backmatter

\bmhead{Acknowledgements}
The authors acknowledge funding from the German Federal Ministry of Education and Research under the funding program ”Foerderprogramm Quantentechnologien – von den Grundlagen zum Markt” (funding program quantum technologies—from basic research to market), project Q-Grid, 13N16177. We also thank Ivan Angelov, Thomas Morstyn, Antonio Macaluso, Supreeth Mysore Venkatesh, and Claudia Linnhoff-Popien for helpful discussions regarding the problem formulations and benchmarking.

\bmhead{Statements and Declarations}
The authors have no competing interests financially or non-financially that are directly or indirectly related to the work submitted for publication. 


\printbibliography

\end{document}